\documentclass[aps,prb,showpacs,reprint,superscriptaddress]{revtex4-1}

\usepackage{graphicx}
\usepackage{color}
\usepackage{amsmath}
\usepackage{bbm}
\usepackage{amssymb}

\usepackage{dsfont}

\usepackage[utf8]{inputenc}	
\usepackage[T1]{fontenc}

\input epsf

\begin{document}

\title{Mixed singlet-triplet superconducting state within the moir\'e $t$-$J$-$U$ model as applied to the description of twisted WSe$_2$ bilayer}
\author{M. Zegrodnik}
\email{michal.zegrodnik@agh.edu.pl}
\affiliation{Academic Centre for Materials and Nanotechnology, AGH University of Krakow, Al. Mickiewicza 30, 30-059 Krakow,
Poland}
\author{A. Biborski}
\affiliation{Academic Centre for Materials and Nanotechnology, AGH University of Krakow, Al. Mickiewicza 30, 30-059 Krakow,
Poland}


\begin{abstract}
We analyze an analog of the $t$-$J$-$U$ model as applied to the description of a single moir\'e flat band of twisted WSe$_2$ bilayer. To take into account the correlation effects induced by a significant strength of the Coulomb repulsion, we use the Gutzwiller approach and compare it with the results obtained by the Hartree-Fock method. We discuss in detail the graduate appearance of a two dome structure of the superconducting state in the phase diagram by systematically increasing the Coulomb repulsion integral, $U$. The two superconducting domes residing on both sides of a Mott insulating state can be reproduced for a realistic parameter range in agreement with the available experimental data. According to our analysis the paired state has a highly unconventional character with a mixed $d+id$ (singlet) and $p-ip$ (triplet) symmetry. Both components of the mixed paired state are of comparable amplitudes. However, as shown here, a transition between pure singlet and pure triplet pairing should be possible in the considered system by tuning the top and bottom gate voltage, which controls the magnitude of valley-dependent spin-splitting in the system. 
\end{abstract}

\maketitle

\section{Introduction}
In recent years the twisted moir\'e structures have gathered a significant amount of interest due to their rich physics on one side and high degree of tunability on the other. The development in the field has been mostly triggered by the discovery of both correlated insulating state and unconventional superconductivity in the magic angle twisted bilayer graphene (TBG)\cite{Cao2018_1, Cao2018_2}. Since than, many other moir\'e structures have been intensively studied which are bilayer boron nitride\cite{Xian2019}, multilayer graphene\cite{Zhang_2021_prl, Chen2019, Shen2020} as well as both homo- and hetero-bilayer transition metal dichalcogenides (TMD)\cite{Wang2020, Liao2020, Ulstrup2020, Ghiotto2021}.

In moir\'e systems, for certain twist angles between the atomic layers, flat energy bands are created. This feature points to a significant role of inter-electronic interactions which is a situation similar to that known from the so-called strongly correlated electron systems such as copper based high-temperature superconductors\cite{Spalek2022}. In fact, both the well known cuprates and the novel moir\'e systems exhibit similar feature in the phase diagram which is the appearance of superconducting (SC) domes residing in close proximity of the correlated insulating state\cite{Cao2018_2,Wang2020}. 

For the case of TBG the flat bands and the resulting correlation induced phenomena are fragile when it comes to deviations of the twist angle around the so-called magic values. On the other hand, in TMD moir\'e systems the effect is more robust and the flat bands are observed for larger spectrum of angles\cite{Wang2020, An2020, Ghiotto2021}. What also distinguisches the TMD systems from those graphene based is the appearance of significant spin-orbit coupling.

As it has been shown both by the DFT calculations and continuum model based approach, the twisted homo-bilayer WSe$_2$ (tWSe$_2$) can be described by a single-band tight-binding model on a triangular lattice with spin-orbit coupled hoppings which reproduce the flat electronic band with a valley-dependant spin-splitting\cite{Wang2020,Haining2020}. In such an approach the so-called displacement field controlled by the bias voltage across the bilayer, tunes the spin-dependent phase of the hopping amplitudes. This allows for changing the magnitude of the spin-splitting in the system. In order to take into account the electron-electron interactions, such approach can be supplemented with the onsite Coulomb repulsion term leading to the Hubbard model. It has been estimated that the value of the Coulomb integral $U$ places tWSe$_2$ in the moderate correlations regime\cite{Wang2020}. However, the magnitude of electronic correlations in those systems can be controlled by the twist angle which changes the $U/W$ parameter, with $W$ being the bare band width. The Hubbard model has been studied in the context of both magnetically ordered states and $d+id$ spin-singlet supercondcutivity of tWSe$_2$\cite{Zang2021,Belanger2022,Zang2022,Haining2020,Haining2022}. In such picture the SC pairing comes as a result of the electronic correlations which become significant for large enough value of the Hubbard $U$. Another canonical model which can be analyzed in the context of electron-electron correlation effects in TMD moir\'e bands is the $t$-$J$ model\cite{Wu2018,Huang2022,Chen2023} within which the Cooper pairing comes naturally from the exchange interaction term $\sim J$. However, it should be noted that the situation in this context is not completely resolved so far and a well established description of the paired state in those systems has not been reached. 

Here, we propose an analog of the $t$-$J$-$U$ model as applied to the description of a single moir\'e flat band of tWSe$_2$. The $t$-$J$-$U$ hamiltonian has been initially proposed for the description of copper based high-T$_C$ superconductors\cite{Daul2000,Basu2001,Xiang2009} and as shown by us in previous years it leads to good quantitative agreement between theory and experiment for selected fundamental features of the paired state\cite{Zegrodnik2017}. This fact together with similarities between the cuprates and twisted moir\'e van der Waals structures constitutes the main motivation for the analysis of unconventional superconducting phase in tWSe$_2$ with the use of the $t$-$J$-$U$ based approach. Due to the significant magnitude of the onsite Coulomb repulsion in such an analysis one has to take into account the electronic correlation effects beyond the level of a simple Hartree-Fock method (HF). Here, we use the Gutzwiller approach (GA), which is efficient and catches the basic effects related with electron-electron interactions. As we show the two superconducting domes residing on both sides of the insulating state can be reproduced with the use of the presented method in agreement with the experimental indicators\cite{Wang2020}. The character of the obtained superconducting state is highly unconventional with a mixed singlet-triplet Cooper pairing. Interestingly, according to our analysis the balance between the singlet and triplet contributions to the pairing could be controlled by the gate voltage in a real experimental setup. In particular, in proper gate voltage range, a transition between pure singlet and pure triplet pairing can be achieved within a single non-hybrid system.


\section{Model and method}
\begin{figure}[!t]
 \centering
 \includegraphics[width=0.5\textwidth]{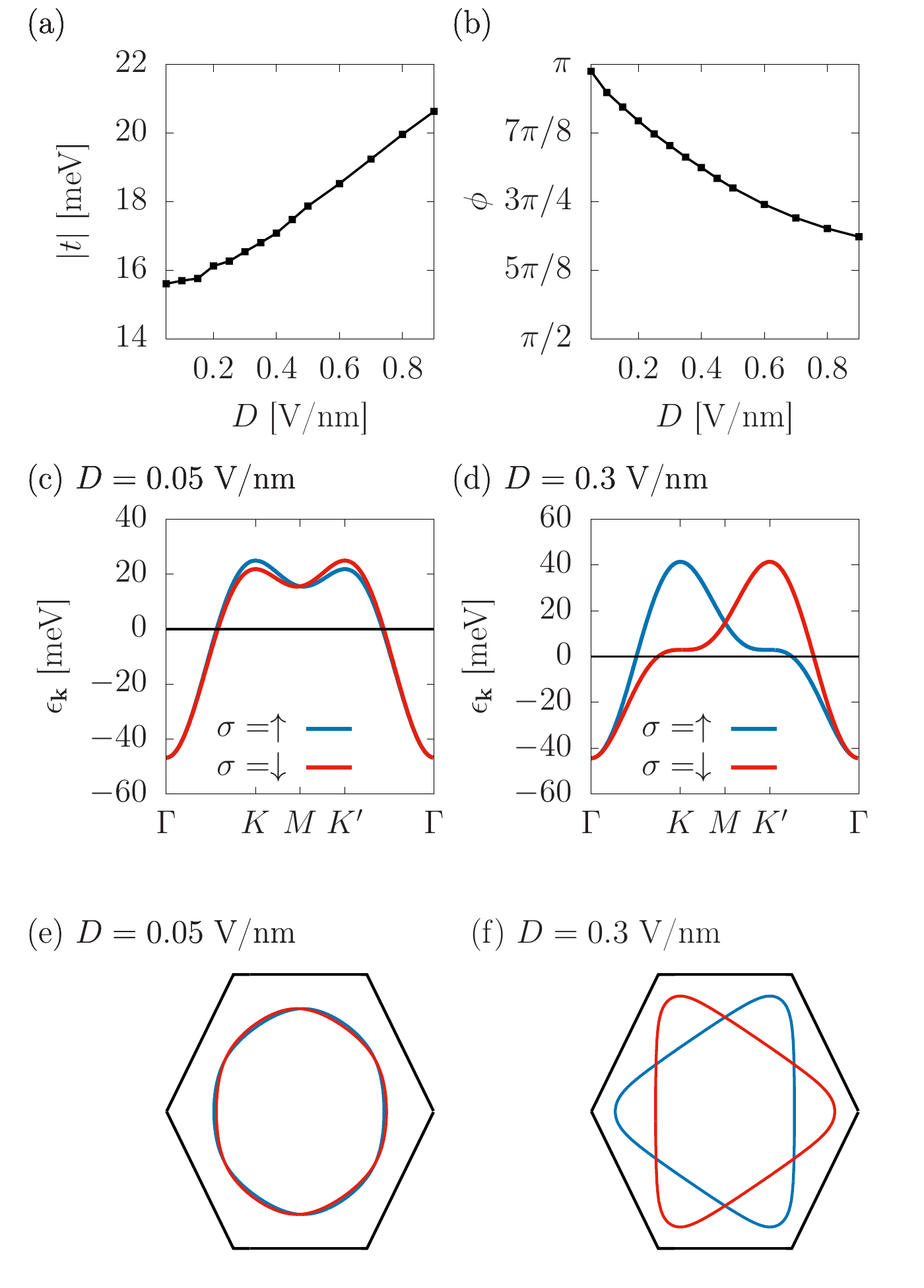}
 \caption{The amplitude (a) and phase (c) of the nearest neighbor complex hopping parameter ($t_{\uparrow}=|t|e^{i\phi}$) as a function of the displacement field as calculated from the data provided in Ref. \onlinecite{Wang2020}. The dispersion relations [(c) and (d)] and the Fermi surfaces at half-filling [(e) and (f)] for two selected values of the displacement field provided in the Figures. Note that by increasing the displacement field one also increases the phase of the hopping parameters which in turn enhances the spin-splitting of the Fermi surfaces.} 
 \label{fig:electronic_structure}
\end{figure}

Here we apply a moir\'e band $t$-$J$-$U$ hamiltonian of the form
\begin{equation}
 \hat{H}=\hat{H}_0+\hat{H}_{J}+\hat{H}_{U}, 
 \label{eq:Hamiltonian_start}
\end{equation}
where $\hat{H}_0$ represents a tight-binding model with spin-orbit coupled hoppings on a triangular lattice. As shown in Refs. \onlinecite{Wang2020,Haining2020}, such single particle hamiltonian can well describe the top moir\'e band of tWSe$_2$. Namely,
\begin{equation}
    \hat{H}_0=\sum_{<ij>\sigma}t_{ij\sigma}\;\hat{c}^{\dagger}_{i\sigma}\hat{c}_{j\sigma},
\end{equation}
where $t_{ij\sigma}$ are the hopping parameters between the nearest-neighboring sites of the form, 
\begin{equation}
    t_{ij\sigma}=|t|e^{i\sigma\phi_{ij}},
    \label{eq:hoppings}
\end{equation}
where $\phi_{ij}=-\phi_{ji}$ due to hermiticity condition and $\sigma=+1,-1$ represents the spin up and down states associated with $+K$ and $-K$ valleys, respectively. The hopping parameters fulfill the $C_3$ rotational symmetry and reconstruct the spin-splitted Fermi surfaces. This is a manifestation of the fact that the system is non-centrosymmetric and spin-orbit coupling appears. In order to simplify the situation, here we limit to the nearest-neighbor hopping terms (summation over $<ij>$) what is justified by the fact that the longer-ranged hoppings are at least one order of magnitude smaller\cite{Wang2020,Haining2020}. In the further analysis we take the values of hoppings provided in Ref. \onlinecite{Wang2020} for the twist angle $5.08^\circ$ which is close to the one for which the signatures of superconductivity have been identified experimentally. As shown in Fig. \ref{fig:electronic_structure}, by changing the value of the so-called displacement field one can tune both $|t|$ and $\phi$, where the latter determines the magnitude of the valley-dependent spin-splitting. In real experimental setup the displacement field is controlled by the bias voltage across the bilayer.

The intersite exchange interaction term appearing in Hamiltonian (\ref{eq:Hamiltonian_start}) has the following form
\begin{equation}
\begin{split}
    \hat{H}_J &= J\sideset{}{'}\sum_{<ij>}\bigg(\hat{S}^z_i\hat{S}^z_j +\cos{(2\phi_{ij})}\sum_{\alpha=x,y}\hat{S}^{\alpha}_i\hat{S}^{\alpha}_j\\
    &+\sin{(2\phi_{ij})}(\mathbf{\hat{S}}_i\times\mathbf{\hat{S}}_j)\cdot \hat{z} \bigg),
\end{split}
\label{eq:J_term}
\end{equation}
where $\hat{S}^{\alpha}_i$ for $\alpha=x,y,z$ are the three components of the electron spin operator and $\phi_{ij}$ is the phase of the spin up hopping between $i$ and $j$ lattice sites. The summation above runs over the nearest neighbor sites only. The prime indicates that each bond between $i$ and $j$ is taken only once. Due to the appearance of the complex spin-dependant hoppings, the exchange term is slightly modified with respect to the one originally appearing within the $t$-$J$-$U$ model as applied for the cuprates\cite{Zegrodnik2017}. In particular, here the Dzyaloshinskii-Moriya (DM) term appears which, however, might be significantly suppressed for very low values of the displacement fields for which $\phi\approx 0$. In such a case the exchange interaction takes the standard form, $\sum_{ij}\mathbf{\hat{S}}_i\mathbf{\hat{S}}_j$.


The last term in Hamiltonian (\ref{eq:Hamiltonian_start}) corresponds to the onsite Coulomb interaction
\begin{equation}
\hat{H}_U=U\sum_{i}\hat{n}_{i\uparrow}\hat{n}_{i\downarrow},
\end{equation}
where $\hat{n}_{i\uparrow}$ ($\hat{n}_{i\downarrow}$) are the number of particle operators on given site with spin up (down), while $U$ corresponds to the Coulomb interaction strength.

In order to take into account the correlation effects induced by a significant value of the Hubbard $U$ we apply the zeroth order diagrammatic expansion of the Gutzwiller wave function (DE-GWF)\cite{Spalek2022} which is equivalent to the $Statistically$ $consistant$ $Gutzwiller$ $Approximation$ (SGA)\cite{Abram2013}. As we show here, for the considered problem the zeroth order expansion provides a good enough approximation to reach a qualitative agreement with fundamental experimental observations for tWSe$_2$. Within this method the many-body effects are captured by taking the variational correlated wave function in the following form
\begin{equation}
    |\Psi_G\rangle=\hat{P}|\Psi_0\rangle,
    \label{eq:correlated_state}
\end{equation}
where $|\Psi_0\rangle$ is the non-correlated (mean-field) state to be defined below and $\hat{P}$ is the Gutzwiller-like correlation operator of the following form
\begin{equation}
    \hat{P}=\prod_i\sum_{\Gamma}\lambda_{i,\Gamma}|\Gamma\rangle_{i\;i}\langle\Gamma|,
\end{equation}
where $|\Gamma\rangle$ are the four states from the local basis $|\emptyset\rangle_i$, $|\uparrow\rangle_i$, $|\downarrow\rangle_i$, $|\uparrow\downarrow\rangle_i$ and $\lambda_{i,\Gamma}$ are the corresponding variational parameters. The $|\Psi\rangle_0$ state is taken here as the uncorrelated superconducting state which is defined by the pairing and hopping mean-field parameters, as well as the electron occupation expectation value
\begin{equation}
\begin{split}
    S_{ij}^{\sigma\bar{\sigma}}=\langle\hat{c}_{i\sigma}\hat{c}_{j\bar{\sigma}}&\rangle_0,\quad P_{ij\sigma}=\langle\hat{c}^{\dagger}_{i\sigma}\hat{c}_{j\sigma}\rangle_0,\\
    &n_{i\sigma}=\langle\hat{n}_{i\sigma}\rangle_0,
\end{split}
\end{equation}
respectively. In the above $\langle\hat{o}\rangle_0$ stands for the expectation value of the $\hat{o}$ operator in the state $|\psi\rangle_0$, and $\bar{\sigma}$ represents spin orientation opposite to $\sigma$.

Here we focus on the unconventional features of the paired state, therefore, for simplicity we assume no magnetic or charge ordering. Hence, we take $\lambda_{i,\Gamma}\equiv \lambda_{\Gamma}$, $\lambda_{\uparrow}=\lambda_{\downarrow}=\lambda_{s}$, $n_{i\uparrow}=n_{i\downarrow}\equiv n_s$. Due to the appearance of the spin-orbit coupling in the model both singlet and triplet pairings are possible which means that we need to treat $S_{ij}^{\uparrow\downarrow}$ and $S_{ij}^{\downarrow\uparrow}$ separately allowing both for pure spin-singlet pairing ($S_{ij}^{\uparrow\downarrow}=-S_{ij}^{\downarrow\uparrow}$), pure spin-triplet pairing ($S_{ij}^{\uparrow\downarrow}=S_{ij}^{\downarrow\uparrow}$) as well as their mixture. As can be shown the equal spin state (ESP) with $S_{ij}^{\uparrow\uparrow}=S_{ij}^{\downarrow\downarrow}$ and $S_{ij}^{\sigma\bar{\sigma}}=0$ is equivalent to the $m_s=0$ spin triplet state with $S_{ij}^{\uparrow\downarrow}=S_{ij}^{\downarrow\uparrow}$ and $S_{ij}^{\sigma\sigma}=0$ and as such is taken into account in our analysis\cite{Smidman_2017}. 

 
The main task within our approach is to calculate the ground-state energy per lattice site of the system in the correlated state within the zeroth order diagrammatic expansion
\begin{equation}
    E_G=\frac{\langle\Psi_G|\hat{H}|\Psi_G\rangle}{N\langle\Psi_G|\Psi_G\rangle}=\frac{1}{N}\langle\hat{H}\rangle_G,
    \label{eq:ground_state_energy}
\end{equation}
where $N$ is the number of moir\'e lattice sites. As we show in more detail in Appendix A, the ground state energy in the correlated state can be expressed as a function of the variational parameters, $\lambda_{\Gamma}$, the  mean-field parameters, $S_{ij}^{\sigma\bar{\sigma}}$ and $P_{ij\sigma}$, as well as $n_s$. In order to carry out the minimization procedure effectively with the self-consistency condition fulfilled we apply the Effective Hamiltonian Scheme\cite{Kaczmarczyk2015}, which we show in some more detail in Appendix B as applied to the considered situation. 

After determining the mean-field and the variational parameters one can calculate the correlated superconducting gap amplitude as well as the hopping expectation values in the correlated state
\begin{equation}
\begin{split}
    \Delta^{\sigma\bar{\sigma}}_{ij}&=\langle\hat{c}_{i\sigma}\hat{c}_{j\bar{\sigma}}\rangle_G=q^2\langle\hat{c}_{i\sigma}\hat{c}_{j\bar{\sigma}}\rangle_0,\\
    \tilde{P}_{ij\sigma}&=\langle\hat{c}^{\dagger}_{i\sigma}\hat{c}_{j\sigma}\rangle_G=q^2\langle\hat{c}^{\dagger}_{i\sigma}\hat{c}_{j\sigma}\rangle_0,
\end{split}
\label{eq:gap_corr}
\end{equation}
where $q=\lambda_s(\lambda_d\;n_s+\lambda_{\emptyset}(1-n_s))$. Since both the hopping and the exchange interactions are taken into account to the nearest neighbors only, we also calculate the six nearest neighbor real space pairing amplitudes both for $\uparrow\downarrow$ and $\downarrow\uparrow$ spin configurations. We use the following notation
\begin{equation}
    \Delta^{\sigma\bar{\sigma}}_{\theta}=\langle\hat{c}_{i\sigma}\hat{c}_{j(\theta)\bar{\sigma}}\rangle_G,
    \label{eq:gap_theta}
\end{equation}
where $j(\theta)$ is the nearest neighboring lattice site of site $i$, while $\theta$ corresponds to the angle between the positive half-$x$ axis and the $\mathbf{R}_{ij}=\mathbf{R}_i-\mathbf{R}_j$ vector. The angles $\theta\in\{0,\;\pi/3,\;2\pi/3,\;\pi,\;4\pi/3,\;5\pi/3\}$ correspond to six subsequent nearest neighbor lattice sites, $j$. Due to the homogeneity of the system, the $\Delta^{\sigma\bar{\sigma}}_{\theta}$ gap amplitude does not depend on the position of the $i$ lattice site. In order to identify the symmetry of the obtained paired state for a given spin configuration we transform the six nearest neighbor pairing amplitudes into a six symmetry resolved pairing amplitudes of the following form
\begin{equation}
    \Delta^{\sigma\bar{\sigma}}_{M,p}=\frac{i^p}{6}\sum_\theta e^{-iM\theta}\Delta^{\sigma\bar{\sigma}}_{\theta},
    \label{eq:gap_symmetry_resolved}
\end{equation}
where the summation runs over the angles $\theta\in\{0,\;\pi/3,\;2\pi/3,\;\pi,\;4\pi/3,\;5\pi/3\}$, and $M$ is the symmetry factor, which takes integer values and corresponds to six possible pairing symmetries (cf. Table \ref{tab:symmetries}). The $p$ parameter corresponds to the parity of the particular symmetry. Namely, for even parity symmetries we have $p=0$ and for odd parities we have $p=1$.
\begin{table}
 \caption{The six possible values of the symmetry factor, $M$, and the corresponding values of the parity factor $p$ (second column), appearing in Eq. (\ref{eq:gap_symmetry_resolved}). The corresponding six symmetries of the superconducting gap together with their parities are provided in the third and fourth columns. The Cooper pair spin state compatible with a given symmetry is provided in the fifth column.}
\begin{center}
\begin{tabular}{ c|c|c|c|c } 
 \hline\hline
 M & p & gap symmetry & parity & spin state \\ 
 \hline
 0 & 0 & $extended$ $s$ & even & singlet \\ 
 \hline
 1 & 1  & $p_x+i\;p_y$ & odd & triplet \\ 
 \hline
 2 & 0  & $d_{x^2-y^2}+i\;d_{xy}$ & even & singlet \\ 
 \hline
 3 & 1  & $f$ & odd & triplet \\ 
 \hline
 4 & 0  & $d_{x^2-y^2}-i\;d_{xy}$ & even & singlet \\ 
 \hline
 5 & 1  & $p_x-i\;p_y$ & odd & triplet \\ 
 \hline\hline
 \end{tabular}
 \label{tab:symmetries}
\end{center}
\end{table}
Finally, the correlated gap amplitudes corresponding to subsequent symmetries can be transformed to those corresponding to singlet and triplet spin states of the Cooper pairs
\begin{equation}
\begin{split}
    \Delta^{s}_{M,p}&=(\Delta^{\uparrow\downarrow}_{M,p}-\Delta^{\downarrow\uparrow}_{M,p})/\sqrt{2},\\
    \Delta^{t}_{M,p}&=(\Delta^{\uparrow\downarrow}_{M,p}+\Delta^{\downarrow\uparrow}_{M,p})/\sqrt{2}.\\
    \label{eq:gap_symmetries_spin}
\end{split}
\end{equation}
Only three of the considered symmetries are compatible with singlet ($d+id$-, $d-id$-, and $extended$ $s$-$wave$) which means that they always give zero for the triplet component ($\Delta^{t}_{M,p}=0$ for $M=0,2,4$). The remaining $p+ip$-, $p-ip$-, and $f$-$wave$ symmetries are compatible with the spin-triplet pairing, therefore, $\Delta^s_{M,p}=0$ for $M=1,3,5$. In our calculation scheme we determine all the twelve nearest neighbor gap amplitudes, $\Delta^{\uparrow\downarrow}_{\theta}$, $\Delta^{\downarrow\uparrow}_{\theta}$, in the correlated state $|\Psi_G\rangle$ without any constraints which would limit us only to one selected symmetry among those provided in Table \ref{tab:symmetries}. Next, we transform them to the final symmetry resolved singlet and triplet gap amplitudes by using Eqs. (\ref{eq:gap_symmetry_resolved}), (\ref{eq:gap_symmetries_spin}). In such a manner we can encompass both the pure superconducting symmetries as well as their mixtures. For the sake of clarity, in the following Sections, we use the symmetry names ($p\pm ip$, $d\pm id$, $f$ etc.) in the subscripts of the symmetry resolved superconducting gaps instead of the values of the $M$ and $p$ factors.

It should be noted that within the so-called diagrammatic expansion method (DE-GWF) one starts  from  the renormalized mean-field  theory  as  the  zeroth-order  result  and  proceed with incorporating systematically the nonlocal correlations of increased range in higher orders. Therefore, within the higher order calculations the correlated gaps and hopping expectation values are no longer related with their non-correlated counterparts by a simple renormalization given by Eqs. (\ref{eq:gap_corr}) (cf. Ref. \onlinecite{Zegrodnik2017}). Also, in such case the hopping and pairing between longer-range neighboring sites needs to be included which makes the analysis much more involved. This allows to take into account the correlation effects with better accuracy. Nevertheless, previous calculations carried out within the SGA approach (equivalent to the zeroth order DE-GWF) and the higher order DE-GWF method, both has lead to the stability of the $d$-$wave$ paired state, and the dome like behavior of the SC gap for the case of the single band $t$-$J$-$U$ model on the square-lattice\cite{Abram2013,Zegrodnik2017}. Therefore, one would expect that also here the inclusion of the higher order terms would not change the main result of the analysis, which is related to the symmetry of the gap and the general form of the phase diagram.

The code which was written to carry out the numerical calculations as well as the data behind the figures presented in the following Sections are available in the open repository\cite{zegrodnik_michal_2023_7753134}.

\section{Results}
In the following we analyze the features of the unconventional superconducting state for the WSe$_2$ bilayer with the twist angle $5.08^\circ$ and the displacement field $D=0.45$ V/nm for which the signatures of both the correlated insulating state and superconductivity have been identified experimentally\cite{Wang2020}. The bare band width for the corresponding hopping parameters is $W\approx 90$ meV. It has been estimated the onsite Coulomb repulsion integral, $U$, is comparable to $W$ for the considered system\cite{Wang2020,Haining2020}. Unless stated otherwise, we take $U=120$ meV which places the system close to the moderately correlated regime. Within the present approach we take the value of $J$ as determined from the formula $J=4|t|^2/U=2.54$ meV corresponding to the mapping of the Hubbard model to the spin Heisenberg model shown in Ref. \onlinecite{Haining2020}. 
\begin{figure}[!t]
 \centering
 \includegraphics[width=0.5\textwidth]{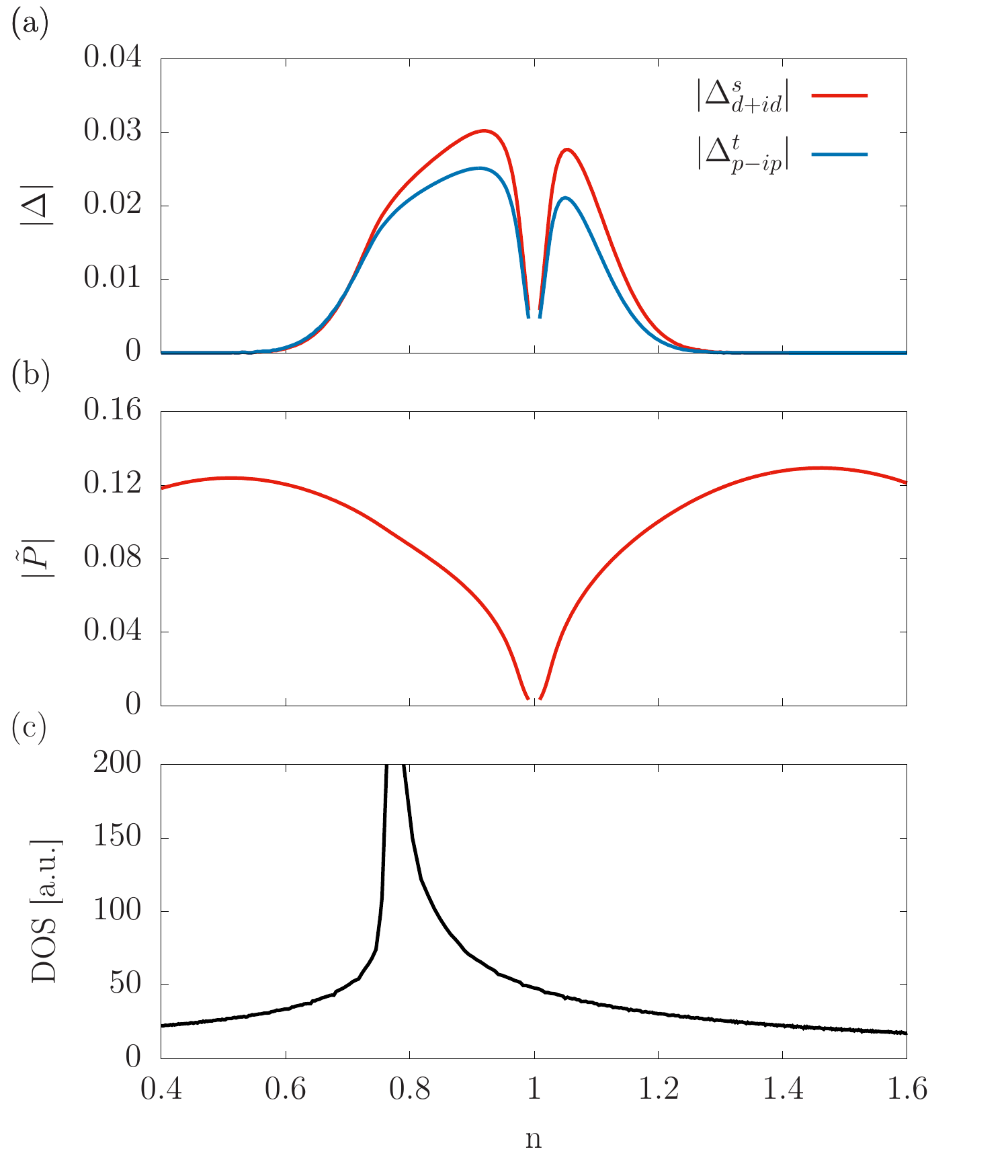}
 \caption{(a) Superconducting gap amplitudes for the $d+id$ (spin singlet) and $p-ip$ (spin triplet) symmetries as a function of band filling $n$; (b) Absolute value of the nearest neighbor hopping expectation value as a function of $n$; (c) Density of states as a function of $n$. The gap amplitudes corresponding to all four remaining symmetries (cf. Table \ref{tab:symmetries}) have turned out to be zero in the considered band filling range. The interaction parameters of the model are set to $U=120$ meV, $J=2.54$ meV.} 
 \label{fig:SCgap_P_DOS_ndep}
\end{figure}

As one can see in Fig. \ref{fig:SCgap_P_DOS_ndep}, for the chosen model parameters the symmetry resolved gap amplitudes create two superconducting domes residing on both sides of the half-filled situation ($n=1$). Here, we define the band filling as the number of electrons in the narrow band per single moir\'e lattice site, $n=2n_s$. According to our calculations the paired state has a mixed singlet-triplet character with comparable magnitudes of both components of the pairing. The singlet (triplet) pairing component is realized by the $d+id$ ($p-ip$) symmetry of the superconducting gap. Apart from the two gap amplitudes shown in the Figure, all the remaining ones listed in Table \ref{tab:symmetries} have turned out to be zero for the given range of model parameters. One should note that pairing in the mixed $d/p$ channel has been proposed to appear due to density waves fluctuations very recently according to functional renormalization group calculations\cite{Klebl2022}. Also, the singlet/triplet pairing has been analyzed within the $t$-$J$ model showing the dominant singlet $d$-wave contribution\cite{Chen2023}. The singlet-triplet mixing obtained here is a consequence of spin-orbit coupling, due to which parity is no longer a good quantum number\cite{Smidman_2017}. Among the possible pairing symmetries the $d\pm id$ and $p\pm ip$ optimize the condensation energy due to the fact that they provide a fully gapped situation in contradiction to the $extended$ $s$-$wave$ and $f$-$wave$ symmetries, for which the line nodes would cross the Fermi surface, in the considered doping range (cf. Appendix C). 

In Fig. \ref{fig:SCgap_P_DOS_ndep} (b) we show that the nearest neighbor hopping expectation value $\tilde{P}$ is being significantly suppressed as one approaches the half-filled situation ($n=1$) from both sides, which indicates the appearance of the correlation induced insulating state (Mott insulator). The insulating behavior suppresses superconductivity in the close proximity of $n=1$ creating the double-dome structure seen in experiments\cite{Wang2020}. The resulting two SC domes are not symmetric with respect to half-filling, which is caused by the appearance of the van Hove singularity for $n\approx 0.8$, which in turn enhances the pairing at the hole-doped side of the phase diagram [cf. Fig. \ref{fig:SCgap_P_DOS_ndep} (c)]. One should note that the approach based on the onsite Gutzwiller-like correlator results in a trivial character of the Mott transition. For the complete description of this effect one should take into account more involved Jastrov and/or doublon-holon correlators\cite{Becca2005,Miyagawa2011}.

\begin{figure}[!h]
 \centering
 \includegraphics[width=0.5\textwidth]{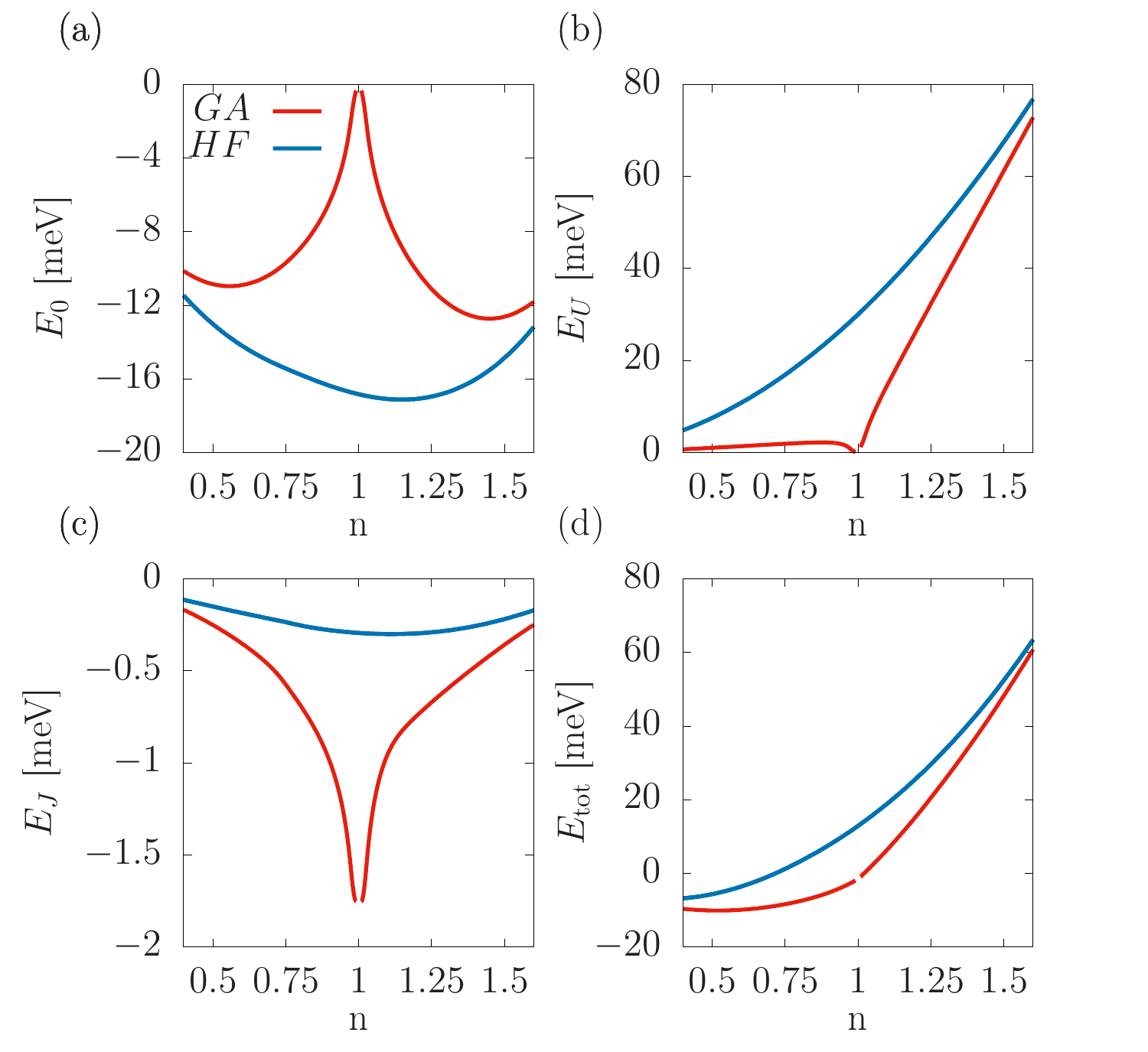}
 \caption{The single particle energy $\langle\hat{H}_0\rangle/N$ (a), the Coulomb interaction energy $\langle\hat{H}_U\rangle/N$ (b), the exchange interaction energy $\langle\hat{H}_J\rangle/N$ (c), and the total energy of the system $\langle\hat{H}\rangle/N$ (d) all as functions of band filling $n$ for the case of Hartree-Fock and Gutzwiller approach calculations. The results have been obtained for the same model parameters as those corresponding to Fig. \ref{fig:SCgap_P_DOS_ndep}.}
 \label{fig:energies_SGA_HF}
\end{figure}

In Fig. \ref{fig:energies_SGA_HF} we show the subsequent contributions to the total system energy as well as their sum, all as functions of band filling as obtained from the Gutzwiller approach and the Hartree-Fock method. The former takes into account the correlation effects induced by the Hubbard $U$ while the latter neglects them. As one can see the two methods deviate significantly close to half-filling where the correlations are most pronounced. Again, within the Gutzwiller approach one can see the signature of the insulating behaviour for $n=1$ where the kinetic energy approaches zero, together with the Coulomb energy as both the electron hopping and the double occupancies are being suppressed. It should be noted that the appearance of the insulating state is concomitant with a significant decrease of the exchange interaction energy as shown in Fig. \ref{fig:energies_SGA_HF} (b). This effect is caused by the fact that the $\lambda_s$ parameter, over which the expectation value of the $\sim J$ term is premultiplied [cf. Eq. (\ref{eq:HJ_corr})], increases close to half filling as shown in Fig. \ref{fig:variational_parameters} (a) in Appendix A. One may understand this as an electron correlation induced increase of the renormalized exchange interaction integral $J'=\lambda_s^4\;J$. As shown in Fig. \ref{fig:energies_SGA_HF} (d), the Gutzwiller method gives total energy lower than HF as it should be since the former is of better quality. 

\begin{figure}[!h]
 \centering
 \includegraphics[width=0.5\textwidth]{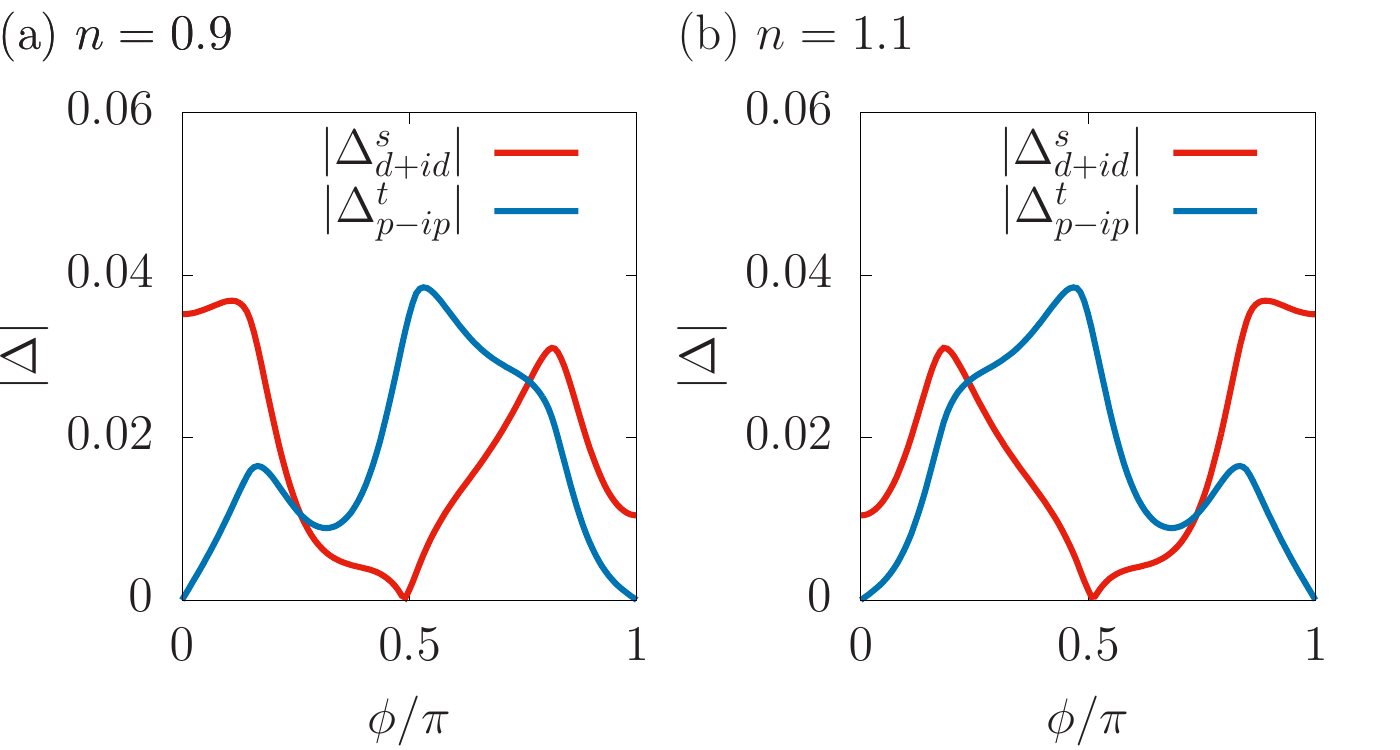}
 \caption{(a) Superconducting gap amplitudes for the $d+id$ (spin singlet) and $p-ip$ (spin triplet) symmetries as a function of the phase of the hopping parameters (cf. Eq. \ref{eq:hoppings}). For the sake of simplicity here we keep the hopping absolute value, $|t|$, constant. (a) and (b) correspond to two different band fillings specified in the Figure.}
 \label{fig:singlet_triplet_phi}
\end{figure}

The appearance of the mixed singlet-triplet pairing seen in Fig. \ref{fig:SCgap_P_DOS_ndep} is a consequence of the spin-orbit coupled hoppings and the resulting spin-splitting of the Fermi surface. It should be noted that with the use of the displacement field one can tune the phase of the hoppings, $\phi$ [cf. Fig. \ref{fig:electronic_structure} (b)], which in turn determines the magnitude of the spin splitting as well as the strength of the DM interaction [cf Eq. \ref{eq:J_term}]. The influence of the DM interaction on the paired state is briefly discussed in Appendix D. Since the spin-splitting plays an important role when it comes to the appearance of the mixed spin-triplet pairing, it is tempting to check if the balance between the singlet and triplet contributions can be modified by changing $\phi$. In Fig. \ref{fig:singlet_triplet_phi} we show the $\phi$-dependance of the two SC gaps, $\Delta^s_{d+id}$ (singlet) and $\Delta^t_{p-ip}$ (triplet), for two selected band fillings. For the sake of simplicity in these calculations we keep $|t|$ constant and change only the phase of the hopping. As one can see by changing $\phi=\pi\rightarrow\pi/2\rightarrow 0$, we also carry out the transition between singlet$\rightarrow$triplet$\rightarrow$singlet pairing, respectively. Purely real (imaginary) hopping values result in purely singlet (triplet) pairing. In between those three distinguished situations a mixed singlet-triplet state appears. The presented data indicate the possibility of tuning the balance between the singlet and triplet contributions by changing the displacement field. The important question is if it is possible to tune the gate voltage in a real experimental setup so as to reach the displacement field values for which $\phi=\pi/2$ in order to generate pure spin-triplet pairing. This would allow for switching between singlet and triplet pairing in an $in$ $situ$ manner by simply adjusting the top and bottom gate voltage. 

\begin{figure}[!t]
 \centering
 \includegraphics[width=0.5\textwidth]{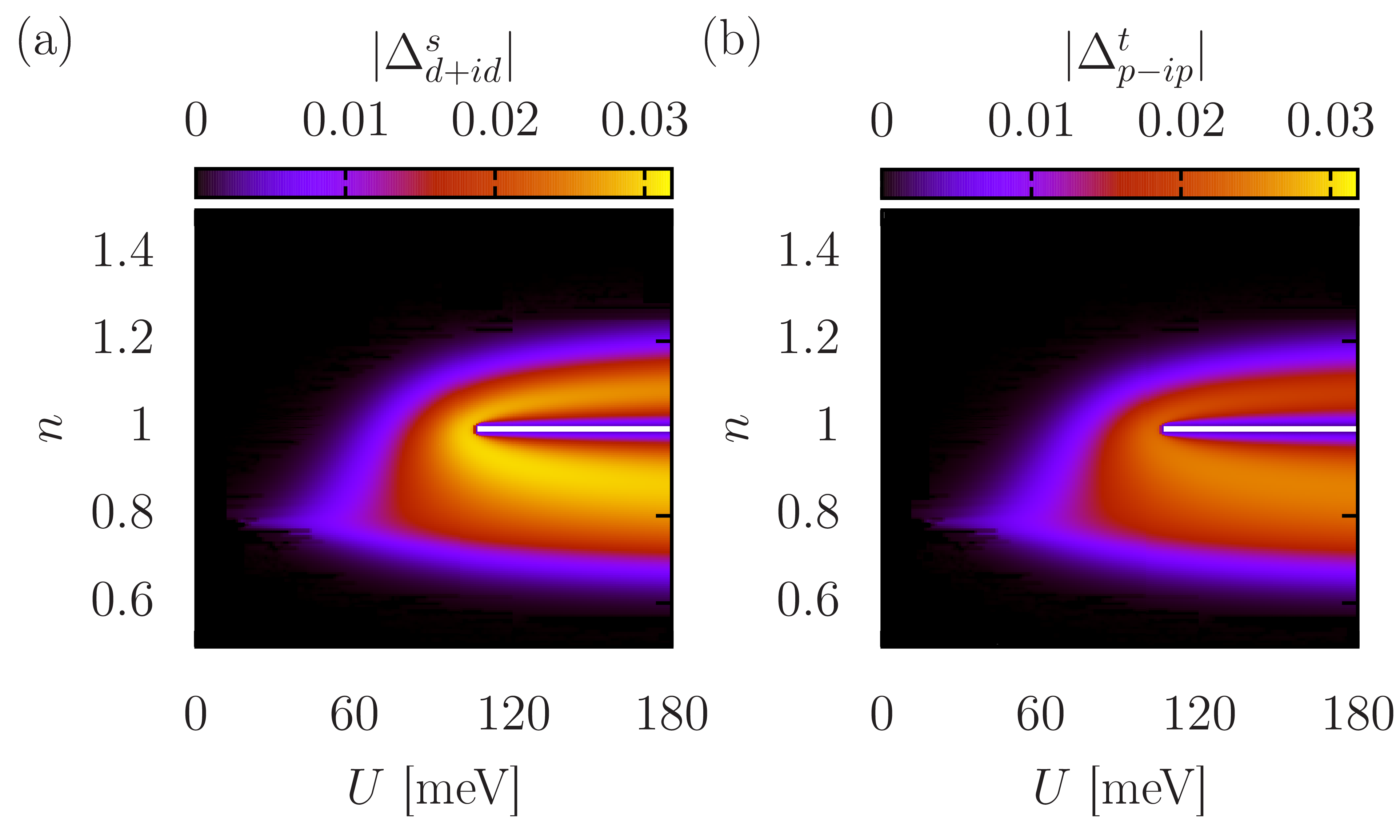}
 \caption{(a) Correlated superconducting gap amplitudes for the $d+id$ (spin singlet) and $p-ip$ (spin triplet) symmetries as a function the Coulomb repulsion integral, $U$, and the band filling $n$ for $J=2.54$ meV. For the chosen model parameters the bare band width is $W\approx 90$ meV. Note that for $U\gtrsim W$ the two SC domes are created in the phase diagram. The narrow white region is where problems with convergence of the numerical procedure appeared.}
 \label{fig:nU_phase_diag}
\end{figure}

As the last part of our analysis we consider the calculated phase diagram in the $(U,n)$ plane. It can be seen from Fig. \ref{fig:nU_phase_diag}, that large enough $U$ is crucial in order to induce the paired state in the system. As the $U$ value is increased, first, a weak superconducting state appears in the range of band fillings which correspond to the van Hove singularity ($n\approx 0.8$) where the high values of density of states create favorable conditions for the pairing. When $U$ approaches the value of $W\approx 90$ meV, the pairing is strengthen around half-filling. This effect is caused by the exchange interaction enhancement which is due to $\lambda_s> 1$, as discussed earlier (cf. Eq. \ref{eq:HJ_corr}). For $U\gtrsim W$, the SC state stability range becomes divided and two SC domes are created which is a direct result of the insulating state appearance at $n=1$. It should be noted that even though the electronic structure of the model and the features of the superconducting state obtained here are significantly different from those corresponding the copper based superconductors there are some clear similarities between the two systems. Namely, in both systems the two superconducting domes divided by a correlated insulating state appear. With this respect, the structure of the $(U,n)$-phase diagram calculated here resembles the one obtained by us previously for the case of a single-band model as applied for the description of the cuprates\cite{Zegrodnik2017,Zegrodnik_2017_2}. However, it should be noted that also other approaches have been reported in recent years which represent different methods and points of view to the problem of pairing in single band models of copper based superconductors\cite{Gunnarsson2015,Qiu2020,Dong2022_1,Dong2022_2}. In particular, the approach presented here does not take into account the dynamical screening effects which leads to frequency dependent $U$ parameter and may affect the strength of the pairing as well as the resulting critical temperature\cite{Nilsson2019}.

\section{Conclusions}
We have applied the moir\'e $t$-$J$-$U$ model to the description of unconventional superconducting state in tWSe$_2$. In order to take into account the electron-electron correlations we have used the Gutzwiller method. As we show our approach leads to the reconstruction of two superconducting domes residing on both sides of the correlation induced insulating state appearing at half-filling, which is in agreement with the experimental data shown in Ref. \onlinecite{Wang2020}. A similar structure of the phase diagram as the one obtained here has also been discussed by us recently in a quantum-dot square lattice system with the use of the variational Monte Carlo method\cite{Biborski2021}. According to our analysis the suppression of the electron hopping processes near the half-filled situation is concomitant with the enhanced role of the exchange interaction term as the corresponding renormalization factor becomes larger than unity, $\lambda_s>1$ (cf. Fig. \ref{fig:variational_parameters} and Eq. \ref{eq:HJ_corr}). This effect also enhances the superconducting phase around the $n=1$ case, however, for large enough value of the Coulomb repulsion integral, $U\gtrsim W$, the appearance of the Mott insulating behaviour separates the SC state stability regime into two superconduting domes (cf. Fig. \ref{fig:nU_phase_diag}). It is worth mentioning that it was proposed very recently within a two-orbital model that the Mott-Hubbard behavior in tWSe$_2$ can be switched into Hund correlated state by tuning the twist angle\cite{Siheon2023}. Such effect is not taken into account here as it requires a two-band extension of the considered model.

The character of the paired state obtained here is highly unconventional with a mixed spin singlet-spin triplet pairing and non-zero values of the the corresponding $d+id$ and $p-ip$ components of the gap symmetry, respectively. As we show the balance between the singlet and triplet contributions can be tuned by adjusting the gate voltage which determines the valley-dependent spin splitting of the electronic structure (cf. Fig. \ref{fig:singlet_triplet_phi}). In particular, if the hopping phase $\phi=\pi/2$ can be reached experimentally, the transition between pure spin-singlet and pure spin-triplet pairing can be realized in the system. 

It should be noted that various magnetic and charge ordered states have been analyzed by a number of methods within the Hubbard and Heisenberg models as applied to the description of tWSe$_2$\cite{Zang2021,Belanger2022,Zang2022,Haining2020,Haining2022,Biderabg2022,Kennes2022}. 
Furthermore, the pair-density-wave state has been shown to appear in weak out-of-plane Zeeman field\cite{Wu2022arxiv}. It is expected that some of those symmetry-broken states may also appear in the present $t$-$J$-$U$ based approach. In particular, in analogy with the cuprates one would expect that close to half-filling antiferromagnetic phase should become stable. As argued in Refs. \onlinecite{Haining2020,Zang2021} the DM interaction acts as an anisotropy that favors in-plane spin ordering and should select the subset of $120^{\circ}$ antiferromagnetic states. In the considered model the DM interaction is affected by the hopping phase [cf. Eq. \ref{eq:J_term}], which in turn can be tuned by the displacement field. Therefore, for particular range of the displacement fields also a tendency towards ferromagnetism can appear. It would be interesting to analyze the interplay between the magnetic/charge ordering and the mixed superconducting pairing considered here. In particular, the appearance of charge density waves may lead to modulation of the pairing amplitudes in space as has been shown by us in recent years for the analogical approach as applied to cuprates\cite{Zegrodnik_2018_cdw}. Such study requires much more involved calculations and is beyond the scope of the present analysis. We should be see progress along this line soon. 

Finally, the spin-triplet $p\pm ip$ paired state is interesting when it comes to its topological features. Namely, if a gate-defined nanowire structure based on tWSe$_2$ could be realized, the $p\pm ip$ pairing could give rise to Majorana bound states at the edges of such nanowire. What is very attractive in such concept is that all the ingredients required for such effect can appear in a single material. This is in opposition to other considered systems in which subsequent factors like spin-orbit coupling or superconducting pairing are introduced with the use of proximity effect in a hybrid system\cite{Oreg2010,Lutchyn2010}.

\section{Acknowledgement}
Discussions with P. W\'ojcik are gratefully acknowledged. This research was founded by National Science Centre, Poland (NCN) according to decision 2021/42/E/ST3/00128. For the purpose of Open Access, the author has applied a CC-BY public copyright licence to any Author Accepted Manuscript (AAM) version arising from this submission.

\appendix
\section{Expectation value of the $t$-$J$-$U$ Hamiltonian in the correlated state}
\begin{figure}[!h]
 \centering
 \includegraphics[width=0.5\textwidth]{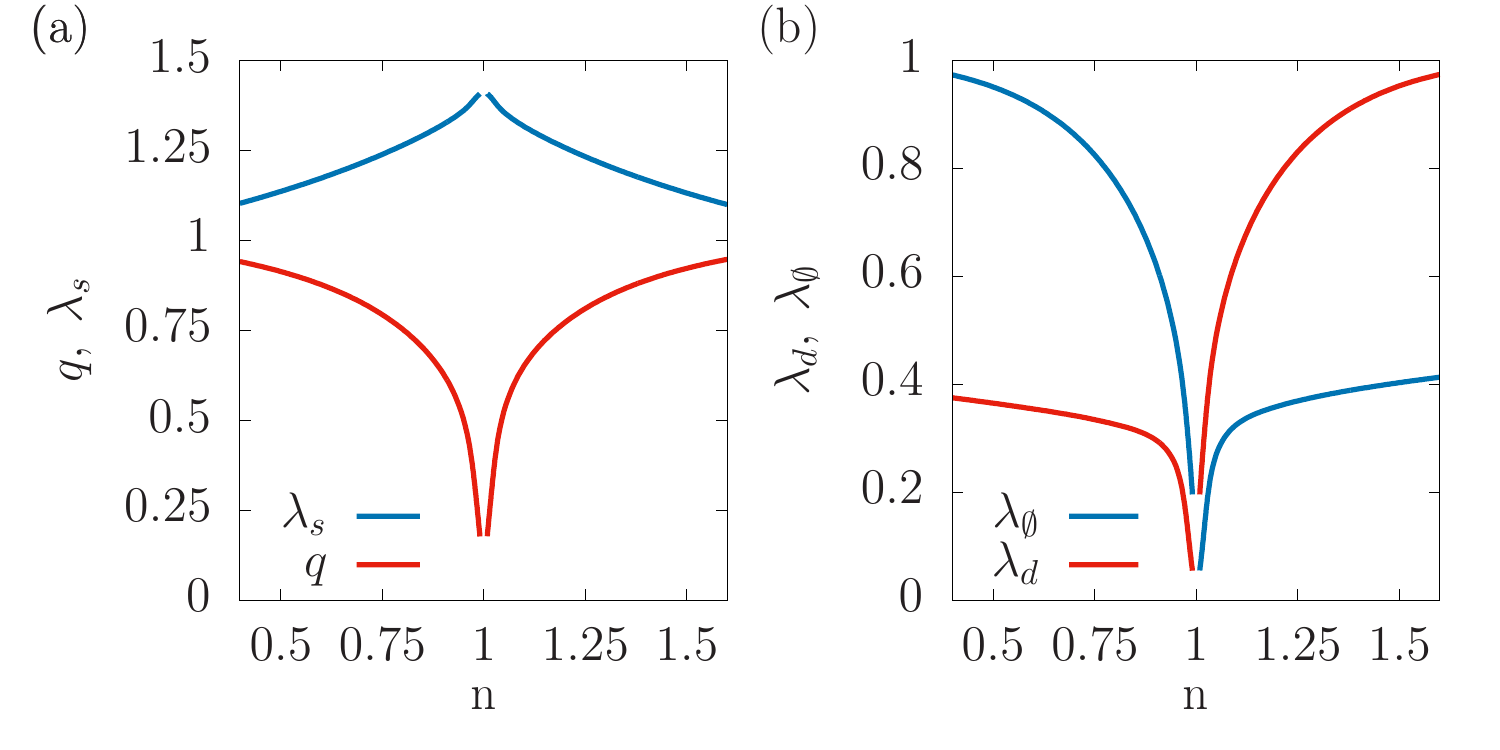}
 \caption{The renoralization factor $q$ and the variational parameters $\lambda_{s}$, $\lambda_{d}$, and $\lambda_{\emptyset}$, as functions of the band filling, $n$, for the same model parameters as those corresponding to Fig. \ref{fig:SCgap_P_DOS_ndep}.}
 \label{fig:variational_parameters}
\end{figure}

Here we show how to express the expectation value of our $t$-$J$-$U$ Hamiltonian in the variational correlated state defined by Eq. (\ref{eq:correlated_state}) within the zeroth order diagrammatic expansion (DE-GWF method). As shown in Ref. \onlinecite{Bunemann_2012}, it is convenient to impose the following condition on the correlation operator
\begin{equation}
    \hat{P}^2_i=1+x\hat{d}_i^{HF},
\end{equation}
where $\hat{d}^{HF}_i=\hat{n}^{HF}_{i\uparrow}\hat{n}^{HF}_{i\downarrow}$, $\hat{n}^{HF}_{i\sigma}=\hat{n}_{i\sigma}-n_s$, and $x$ is yet another variational parameter. One can express the variational parameters $\lambda_{\Gamma}$ as a function of $x$ in the following manner
\begin{equation}
\begin{split}
    \lambda^2_d&=1+x(1-n_s)^2,\\
    \lambda^2_s&=1-xn_s(1-n_s)^2,\\
    \lambda^2_{\emptyset}&=1+xn_s^2,
\end{split}
\end{equation}
therefore, in fact we are left with only one variational parameter. In such case the expectation values of subsequent terms of our Hamiltonian take the form
\begin{equation}
    \langle\hat{H}_0\rangle_G=\sum_{ij}q^2t_{ij\sigma}\langle\hat{c}^{\dagger}_{i\sigma}\hat{c}_{j\sigma}\rangle_0,
    \label{eq:hopping_corr}
\end{equation}

\begin{equation}
\begin{split}
    \langle\hat{H}_J\rangle_G&=\lambda^4_s\;J\sum_{<ij>}\bigg(\frac{1}{2}\;\sum_{\sigma}\;e^{i\sigma2\phi_{ij}}\langle\hat{c}^{\dagger}_{i\sigma}\hat{c}_{i\bar{\sigma}}\hat{c}^{\dagger}_{j\bar{\sigma}}\hat{c}_{j\sigma} \rangle_0\\
    &+\frac{1}{4}\;\sum_{\sigma\sigma'}\sigma\sigma'\langle\hat{n}^{HF}_{i\sigma}\hat{n}^{HF}_{j\sigma'}\rangle_0\bigg),
\end{split}
\label{eq:HJ_corr}
\end{equation}

\begin{equation}
\langle\hat{H}_{U} \rangle=N\lambda_d^2\;U\; n_s^2,    
\label{eq:HU_corr}
\end{equation}
where $q=\lambda_s\big(\lambda_d n_s+\lambda_{\emptyset}(1-n_s)\big)$. Since all the four-operator expectation values which appear on the right-hand side of Eq. (\ref{eq:HJ_corr}) are in the non-correlated state, one can carry out the standard Wick's theorem decomposition in order to express them in terms of the mean-field parameters $S_{ij}^{\sigma\bar{\sigma}}$, $P_{ij\sigma}$. For the sake of clarity in Fig. \ref{fig:variational_parameters} we provide the band filling dependence of the renormalization factor $q$ and the variational parameters $\lambda_{s}$, $\lambda_{d}$, and $\lambda_{\emptyset}$, appearing in expressions (\ref{eq:hopping_corr}-\ref{eq:HU_corr}). It should be noted that the Hartree-Fock calculations are equaivalent to the situation for which $\lambda_{\Gamma}\equiv 1$, $q\equiv 1$.

\section{The effective Hamiltonian scheme}
From the minimization condition of the ground state energy (\ref{eq:ground_state_energy}) one can derive the effective Hamiltonian, which for the case of superconducting phase has the form
\begin{equation}
\begin{split}
 \hat{\mathcal{H}}_{\textrm{eff}}&=\sum_{ij\sigma}t^{\textrm{eff}}_{ij\sigma}\hat{c}^{\dagger}_{i\sigma}\hat{c}_{j\sigma}-\mu^{\textrm{eff}}\sum_{i\sigma}\hat{n}_{i\sigma}\\
 &+\sideset{}{'}\sum_{<ij>\sigma}\big(\Delta^{\textrm{eff}}_{ij\sigma\bar{\sigma}}\hat{c}_{j\sigma}\hat{c}_{i\bar{\sigma}}+H.c.\big),
 \end{split}
 \label{eq:H_effective}
\end{equation}
where the effective hopping, effective chemical potential, and the effective superconducting gap parameters are defined through the corresponding relations
\begin{equation}
 t^{\textrm{eff}}_{ij\sigma}\equiv \frac{\partial\mathcal{F}}{\partial P_{ij\sigma}},\quad (\Delta^{\textrm{eff}}_{ij\sigma\bar{\sigma}})^*\equiv \frac{\partial\mathcal{F}}{\partial S^{\sigma\bar{\sigma}}_{ji}},\quad\mu^{\mathrm{eff}}\equiv-\frac{\partial\mathcal{F}}{\partial n_s},
 \label{eq:effective_param}
\end{equation}
where $\mathcal{F}=E_G-\mu_Gn_s$ with $\mu_G$ being the chemical potential determined in the correlated state. By using Eqs. (\ref{eq:hopping_corr}-\ref{eq:HU_corr}) one can express the ground state energy, $E_G$ in the correlated state with the use of the mean-field parameters, variational parameters, as well as the band filling. This allows to calculate the values of the effective parameters from Eq. (\ref{eq:effective_param}). Next, the effective Hamiltonian can be diagonalyzed by introducing the generalized Bogolubov de Gennes transformation and the self-consistent equations for $P_{ij\sigma}$, $S^{\sigma\bar{\sigma}}_{ij}$, $\mu_G$ can be derived in a standard manner. Additionally, the procedure of solving the self-consistent equations has to be supplemented with the minimization of the ground state energy with respect to the variational parameter $x$.

\section{Gap symmetry in the $\mathbf{k}$-space}
Here we show explicitly the $\mathbf{k}$-dependency of the superconducting gap for selected pairing symmetries. It should be noted that in order to obtain the $\mathbf{k}$-dependant superconducting gaps one can transform the real-space pairing amplitudes given by equations (\ref{eq:gap_theta}) in the following manner
\begin{equation}
    \Delta^{\sigma\sigma'}(\mathbf{k})=\sum_{\theta}\Delta^{\sigma\sigma'}_{\theta}e^{i\mathbf{k}\mathbf{R}_{ij(\theta)}},
\end{equation}
where $j(\theta)$ is the nearest neighboring lattice site of site $i$, while $\theta$ corresponds to the angle between the positive half-$x$ axis and the $\mathbf{R}_{ij}=\mathbf{R}_i-\mathbf{R}_j$ vector. The summation is over the angles $\theta\in\{0,\;\pi/3,\;2\pi/3,\;\pi,\;4\pi/3,\;5\pi/3\}$, which correspond to six subsequent nearest neighbor lattice sites, $j$. In analogy to the real space gap amplitudes [cf. Eq. (\ref{eq:gap_symmetries_spin})], also here we can define the singlet and triplet superconducting gaps in $\mathbf{k}$-space.
\begin{equation}
\begin{split}
    \Delta^{s}(\mathbf{k})&=\big(\Delta^{\uparrow\downarrow}(\mathbf{k})-\Delta^{\downarrow\uparrow}(\mathbf{k})\big)/\sqrt{2}\\
    \Delta^{t}(\mathbf{k})&=\big(\Delta^{\uparrow\downarrow}(\mathbf{k})+\Delta^{\downarrow\uparrow}(\mathbf{k})\big)/\sqrt{2}
\end{split}
\end{equation}
\begin{figure}[!h]
 \centering
 \includegraphics[width=0.5\textwidth]{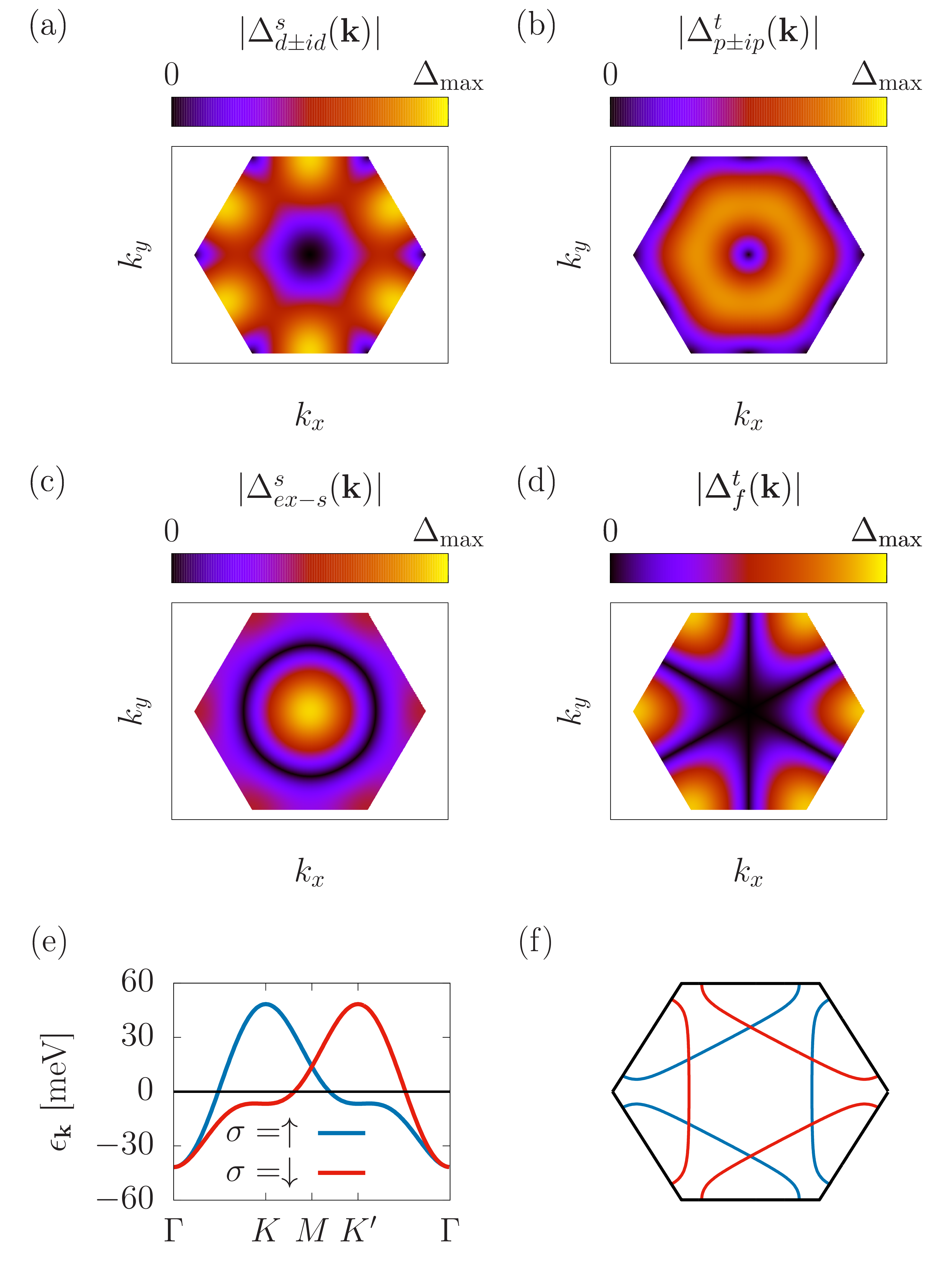}
 \caption{The absolute value of the superconducting gaps for spin-singlet symmetries ($d\pm id$-wave and $extended$ $s$-wave) and spin-triplet symmetries ($p\pm ip$-wave and $f$-wave), all as functions of momentum, $\mathbf{k}$. In (e) and (f) we show the spin up and down dispersion relations and the Fermi surfaces for exemplary value of the band filling $n=0.9$. Note that the $d\pm id$ and $p\pm ip$ pairing symmetries provide significantly larger values of the superconducting gaps at the Fermi surface than the $exteded$ $s$- and $f$-$wave$ pairings.}
 \label{fig:gap_k_space_sym}
\end{figure}

In Fig. \ref{fig:gap_k_space_sym} we show the absolute value of the superconducting gaps for spin-singlet symmetries ($d\pm id$-$wave$ and $extended$ $s$-$wave$) and two spin-triplet symmetries ($p\pm ip$-$wave$ and $f$-$wave$), all as functions of momentum, $\mathbf{k}$. Additionally in Fig. \ref{fig:gap_k_space_sym} (f) we show the Fermi surface for one exemplary band filling value, $n=0.9$, for which the mixture of $d+id$ and $p-ip$ pairing has been shown to be stable in the considered model (cf. Fig. \ref{fig:SCgap_P_DOS_ndep}). 

As one can see from the figures, the $d\pm id$ and $p\pm ip$ pairing symmetries provide a fully gapped situation at the Fermi surface in contradiction to the $extended$ $s$-$wave$ and $f$-$wave$ symmetries. This leads to the optimization of the condensation energy for the case of the $d\pm id$ and $p\pm ip$ mixing.

\section{The effect of DM interaction on the paired state}
\begin{figure}[!t]
 \centering
 \includegraphics[width=0.5\textwidth]{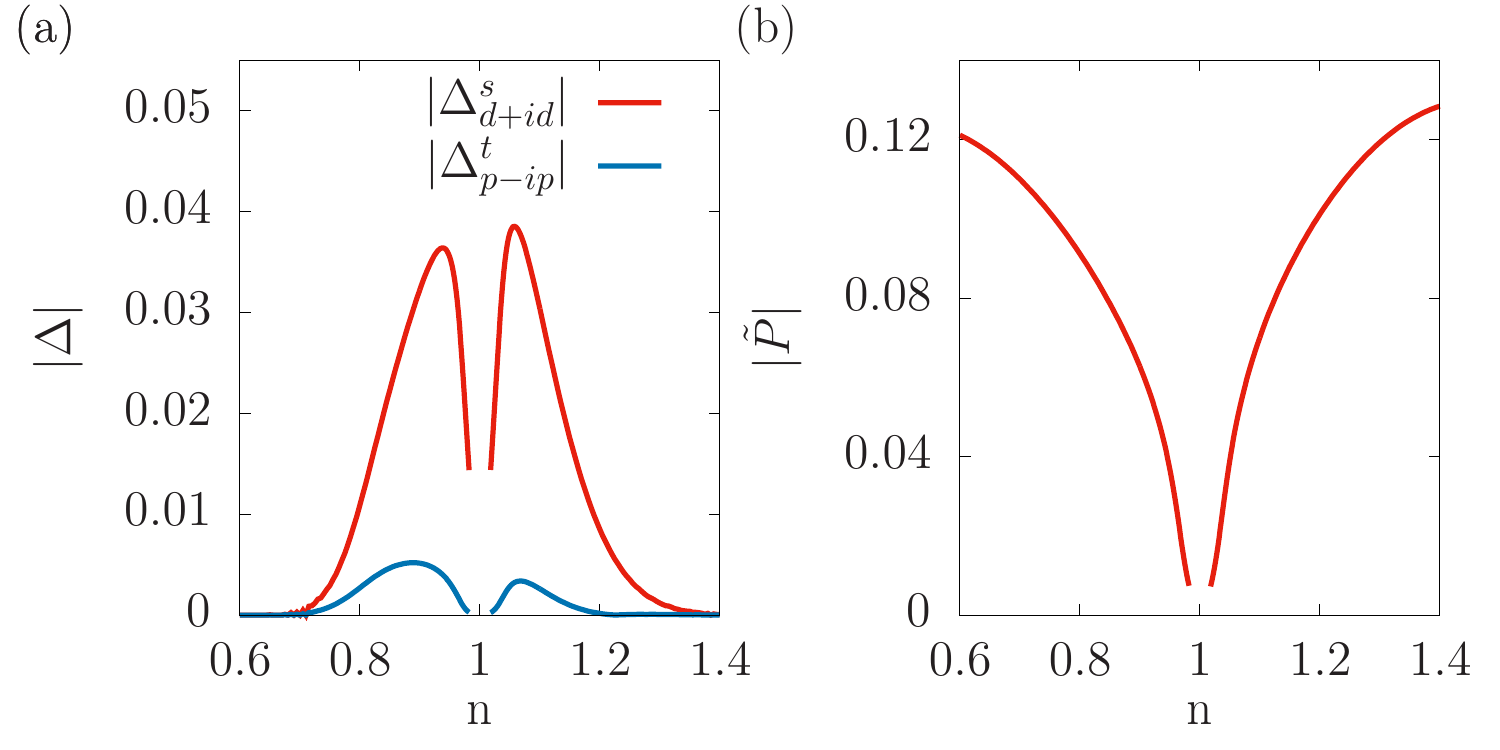}
 \caption{(a) Superconducting gap amplitudes for the $d+id$ (spin singlet) and $p-ip$ (spin triplet) symmetries as a function of band filling $n$; (b) Absolute value of the nearest neighbor hopping amplitude as a function of $n$. Here, the $\sim J$ term is taken in the form given by Eq. \ref{eq:H_J_noDM}. The model parameters are the same as those corresponding to Fig. \ref{fig:SCgap_P_DOS_ndep}.}
 \label{fig:SCgap_P_DOS_ndep_noDM}
\end{figure}

For the sake of completeness, in Fig. \ref{fig:SCgap_P_DOS_ndep_noDM} we show the results obtained for the case when the spin-splitting is still present in the electronic structure, however, the exchange interaction term is taken in the standard form without the DM term. Namely, 
\begin{equation}
    \hat{H}_J=J\sum_{ij}\mathbf{\hat{S}}_i\mathbf{\hat{S}}_j.
    \label{eq:H_J_noDM}
\end{equation}
As one can see in such case the triplet component is still present, however it is significantly smaller with respect to the results presented in Fig. \ref{fig:SCgap_P_DOS_ndep} where the DM term has been taken into account. 

\newpage

\bibliography{refs.bib}

\end{document}